# Absorption and Emission in a Free Electron Lasers

M.A. Kutlan

Institute for Particle & Nuclear Physics, Budapest, Hungary

kutlanma@gmail.com

Multiphoton processes in undulators with plane polarized magnetic field are considered. It is shown that the use of strong magnetic fields in the undulator, for beams with relatively low energy makes it possible to increase substantially the frequencies of the amplified electromagnetic waves without noticeably decreasing the gain. The absorption, emission probabilities and the gain are calculated.

## 1. INTRODUCTION

The operating principle of free-electron lasers is based on the interaction of a beam of relativistic electrons with the stationary periodic magnetic field of an undulator. The development of lasers of this type was reported numerously [1-3]. One of the reasons interactions of the free-electron laser is a possibility of regulating the lasing frequency by varying the electron energy. This also raises hopes of advancing into the ultraviolet and X-rays regions.

At high frequencies, however, the gain obtained by perturbation theory in relatively weak fields of the undulator decreases with increasing frequency $\omega$ of the amplified wave like $\omega^{-2}$, (Ref. [4,5]). Therefore any method of increasing the gain in the infrared, optical, and ultraviolet bands is of interest. In particular, it makes sense to consider multiphoton processes, when the undulator field parameter $K = eH_1\lambda_0/(2\pi mc^2) \geq 1$ ($H_1$ is the amplitude of the undulator magnetic field, $\lambda_0$ is its period, and m is the electron mass).

The saturation of the intensity of the amplified electromagnetic wave in the case of a helical magnetic field of the undulator was considered in [6,7]. In the present paper, the gain is calculated within the framework of a quantum-mechanical description of the behavior of the electrons in specified classical fields. The wave function of the electrons in a strong magnetic field is obtained without using perturbation theory; the field of the amplified wave is regarded as weak, and the probabilities of the induced radiation (absorption) are calculated in first-order perturbation theory with respect to this field. In the derivation of the equations is assumed that the one-electron approximation criterion is satisfied, so that the considered effects are proportional to the first power of the electron density $N_e$ in the beam. For more details on FEL see [8-77].

## 2. BASIC EQUATIONS

We consider the motion of a relativistic electron situ- ated in the spatially periodic magnetic field of an undu- lator and in the field of a traveling electromagnetic wave. We define the 4-potentials of the fields by the equations ($\hbar = c = 1$)

$$A_1(x) = (0, e_x A_1 \cos q_0 z, 0, 0),$$
$$A_2(x) = 1/2 A_2 \left[ e_2 e^{-ikx} + c.c. \right], \quad (1)$$

where $e_1 = (0, e_x, 0, 0)$ and $e_1 = (0, \mathbf{e})$ are the unit vectors the polarization of the fields of the undulator and of the wave, respectively (we are considering the case of planar polarization); $A_1$ and $A_1$ are the corresponding amplitudes; $k_1 = (0, 0, 0, q_0)$ and $k = (\omega, \mathbf{k})$ are the 4-momenta of the undulator field and of the wave respectively, $q_0 = 2\pi / \lambda_0$. We use in (1) the usual notation for the scalar product of 4-vectors: $kx = (kx) = \omega t - \mathbf{k} \cdot \mathbf{r}$.

As the basic equation, neglecting small spin corrections, we use the Klein-Gordon equation in fields $A_{1,2}(x)$. The dimensionless parameter $\zeta = eA_1 / m$, which characterizes the intensity of the interaction of the electron with the undulator field, is assumed to be $\geq 1$ and, accordingly, we take into account the field $A_1(x)$ in all orders of perturbation theory. We assume the field of the wave $A_2(x)$ to be weak enough and consider it in first order of perturbation theory.

The undulator field modulates the $\Psi$ function of the electron in accordance with the equation

$$\left[ -\frac{\partial}{\partial x_\mu} \frac{\partial}{\partial x_\mu} - 2ei \left( A_1^\mu \frac{\partial}{\partial x_\mu} \right) + (eA_1)^2 - m^2 \right] \Psi = 0. \quad (2)$$

The solution of Eq. (2) is given by a function of the type [78]

$$\Psi = e^{-ipx} F(k_1 x) = (2\varepsilon V)^{-1/2} \exp\left[ -i\tilde{p}x - i\frac{e(A_1 p)}{(pk_1)} \sin k_1 x - i \frac{(A_1 p)^2}{8(pk_1)} \sin 2k_1 x \right],$$
$$\tilde{p} = p + \frac{e^2 A_1^2}{4(pk_1)} k_1, \quad (3)$$

Where $\tilde{p}$ is the average kinetic 4-momentum of the electron; $\tilde{p} = (\varepsilon, \mathbf{p})$ is the 4-momentum of the free electron (when the field is turned off). The function (3) corresponds to normalization of the wave function of the free particle to one particle in the volume V.

It should be noted that the function (3), in contrast to the solution of the relativistic equation for a traveling wave [8], is an approximate solution of (2). When (3) was substituted in (2), we left out small terms $\sim eA_1 p_\perp / p_\parallel^2$ and $q_0 / p_\parallel$, where $p_\perp$ and $p_\parallel$ are the components of the electron momentum p perpendicular and parallel to the undulator axis. The solution (3) does not contain the re- flected wave that arises when the particle enters the magnetic field of the undulator. In the case of collinear geometry, which will be considered from now on, the reflected wave can be neglected if the criterion $\zeta^2/\gamma^2 \ll 1$ is satisfied ($\gamma = \varepsilon/m$).

In the case of an helical undulator field the function $(eA_1)^2$ does not contain a spatial dependence, and allowance for the corresponding term in (2) leads only to a renormalization of the electron mass. In the case of a of plane-polarized magnetic field this function contains a periodic dependence on the coordinate, which is in fact the cause of the multiphoton effects in strong fields.

The processes considered are characterized by an S-matrix element given in perturbation theory in the first order of the interaction $\hat{V} = -2e^2(A_1 A_2)$ by

$$S_{ji} = i\frac{e^2 A_1 A_2 (e_1 e_2)}{(\varepsilon \varepsilon')^{1/2} V} \int \exp[i(\tilde{p}' - \tilde{p})x - i\alpha \sin 2k_1 x] \cos kx \, d^4x, \quad (4)$$

where $p' = (\varepsilon', \mathbf{p}')$ is the 4-momentum of the final state of the electron; o! denotes the dimensionless quantity

$$\alpha = \frac{1}{2} e^2 A_1^2 \left[ \frac{1}{(pk_1)} - \frac{1}{(p'k_1)} \right]. \quad (5)$$

We assume that the region of interaction of the electrons with the undulator field is infinite or, more accurately speaking, we assume satisfaction of the condition $\delta\varepsilon/\varepsilon \gg 1/N$, where N is the number of periods of the undulator and $\delta\varepsilon/\varepsilon$ is the relative energy spread in the initial beam of the electrons. In this case the integration in (4) is carried out formally with infinite limits, and the result, as usual, constitutes four $\delta$ functions that yield the system energy and momentum conservation laws.

We thus obtain from (4) the following expressions for the S-matrix elements that describe the processes with emission $(S_{fi}^e)$ and absorption $(S_{fi}^a)$ of a photon of the amplified wave of frequency $\omega$:

$$S_{fi}^{e,a} = i\frac{e^2 A_1 A_2}{4(\varepsilon \varepsilon')^{1/2} V} (2\pi)^4 \sum_n J_n(\alpha) \left\{ \delta^{(4)}\left[\tilde{p}' \pm k - \tilde{p} - (2n-1)k_1\right] + \delta^{(4)}\left[\tilde{p}' \pm k - \tilde{p} - (2n+1)k_1\right] \right\},$$

(6)

where the upper and lower signs correspond respectively to emission and absorption. In the derivation of (6) we used the Fourier expansion

$$\exp(-i\alpha \sin 2k_1 x) = \sum_{n=-\infty}^{+\infty} J_n(\alpha)\exp(-in\, 2k_1 x),$$

where $J_n(\alpha)$ are Bessel functions.

From the conservation laws contained in the $\delta$ functions of Eq. (6)

$$\tilde{p}' \pm k - \tilde{p} - (2n-1)k_1 = 0, \qquad \tilde{p}' \pm k - \tilde{p} - (2n+1)k_1 = 0, \qquad (7)$$

and also from the conditions

$$(\tilde{p})^2 = (\tilde{p}')^2 = \tilde{m}^2 \approx m^2\left(1 + \zeta^2/2\right), \quad k^2 = 0,$$

follow the permissible values of n in the sums of (6). Thus, in the first sum for the process with emission $(S_{fi}^e)$ we obtain n $n \geq 1$, in the second we get $n \geq 0$; for the process with absorption ($(S_{fi}^a)$ we obtain $n \leq 0$ and $n \leq -1$ in the first and second sums respectively. Taking into account the permissible n, expressions (6) can be represented in the form

$$S_{fi}^e = i\frac{e^2 A_1 A_2}{4(\varepsilon\varepsilon')^{1/2} V}(2\pi)^4 \sum_{n=0}^{\infty}\left[J_{n+1}(\alpha)+J_n(\alpha)\right]\delta^{(4)}\left[\tilde{p}' \pm k - \tilde{p} - (2n-1)k_1\right],$$

$$S_{fi}^a = i\frac{e^2 A_1 A_2}{4(\varepsilon\varepsilon')^{1/2} V}(2\pi)^4 \sum_{n=0}^{\infty}(-1)^n\left[-J_{n+1}(\alpha)+J_n(\alpha)\right]\delta^{(4)}\left[\tilde{p}' - k - \tilde{p} - (2n+1)k_1\right] \qquad (8)$$

On going to the probabilities of the processes, the $\delta$ functions of Eqs. (8) determine the three components of the momentum p' of the final state of the electron. The singularity that remains after integration with respect to dp', as is customary in problems of induced radiation in a given field, is eliminated by taking into account the real properties of the interacting objects, such as the spread of the initial electron beam with respect to direction or energy, the finite interaction region, the deviations of the undulator field from periodicity, etc. In our case the total probabilities of the processes should be averaged over the initial energy distribution of the electrons in the beam, given by the distribution function $f(\varepsilon)$. We assume that this function is normalized to unity by the condition

$$\int f(\varepsilon)d\varepsilon = 1$$

and that the half-width of the function $\delta\varepsilon \ll \varepsilon$.

### 3. ABSORBTION AND EMISSION PROBABILITIES AND THE GAIN

Taking all the foregoing into account, the probabilities, per unit time $dw_e$ and $dw_a$ of the processes with emission and absorption of a photon $\omega$ of the amplified wave are given respectively by the expressions (the upper and lower signs in this and all the succeeding equations correspond to emission and absorption, respectively)

$$dw_{e,a} = \frac{\pi}{8}\left(e^2 A_1 A_2\right)^2 \sum_{n=0}^{\infty}\left[J_{n+1}(\alpha_{e,a}) \pm J_n(\alpha_{e,a})\right]^2 \frac{\delta(\varepsilon'_{e,a} \pm \omega - \varepsilon)}{\varepsilon \varepsilon'_{e,a}} f(\varepsilon)d\varepsilon, \qquad (9)$$

where in accordance with the definition (5) and the conservation laws (7)

$$\alpha_{e,a} = \frac{\omega}{1 \pm \omega/\varepsilon}; \qquad \omega = \frac{K^2/2}{1 + K^2/2}\left(n + \frac{1}{2}\right). \qquad (10)$$

From the conservation laws follow also formulas for the frequencies of the emitted we and absorbed w, photons:

$$\omega_{e,a} = \frac{\omega}{1 \pm \omega/\varepsilon}; \qquad \omega = \frac{8\pi\gamma^2}{\lambda_0(1 + K^2/2)}\left(n + \frac{1}{2}\right). \qquad (11)$$

In (9) these are no interference terms corresponding to different n. The results that follow indicate that for the parameters considered in the present paper the frequency difference $\delta\omega \equiv \omega(n + L) - \omega(n)$ is less than the spontaneous emission line width.

The rate of amplification of the wave is determined by the probability difference

$$dw_e - dw_a = \frac{\pi}{8}\left(e^2 A_1 A_2\right)^2 \sum_{n=0}^{\infty}\left\{\begin{array}{l}\left[J_{n+1}(\alpha_e) \pm J_n(\alpha_e)\right]^2 \dfrac{\delta(\varepsilon'_e + \omega - \varepsilon)}{\varepsilon \varepsilon'_e} \\ -\left[J_{n+1}(\alpha_a) \pm J_n(\alpha_a)\right]^2 \dfrac{\delta(\varepsilon'_a \pm \omega - \varepsilon)}{\varepsilon \varepsilon'_a}\end{array}\right\}f(\varepsilon)d\varepsilon. \qquad (12)$$

Averaging over & in (9) and (12) can be easily carried out by using the conditions $q_0 \ll \omega \ll \varepsilon \approx |\mathbf{p}|$. In this approximation, the 6 functions in (9) and (12) can be written in the form

$$\delta\left(\varepsilon'_{e,a} \pm \omega - \varepsilon\right) = \frac{\delta(\varepsilon - \varepsilon_{e,a})}{|\partial \varepsilon'_{e,a}/\partial \varepsilon - 1|} = \frac{\delta(\varepsilon - \varepsilon_{e,a})\varepsilon'_{e,a}}{2(2n+1)q_0}, \qquad (13)$$

where $\varepsilon_{e,a} = \varepsilon_0 \pm \Delta\varepsilon$ are the energies of the electrons that emit or absorb a photon of given frequency $\omega$ for a given period $\lambda_0 = 2\pi/q_0$ of the undulator;

$$\varepsilon_0 = \tilde{m}\left[\frac{\omega}{2q_0}(2n+1)\right]^{1/2}; \qquad \Delta\varepsilon = \frac{\omega}{2}. \qquad (14)$$

After substituting (13) in (12) and integrating with respect to d&, we obtain for the difference between the total probabilities, per unit time, of emission and absorption of a photon of frequency $\omega$,

$$\Delta w \equiv w_e - w_a = \frac{\pi}{8}\left(e^2 A_1 A_2\right)^2 \frac{\left[J_{n_0+1}(\alpha_e) \pm J_{n_0}(\alpha_e)\right]^2}{2n_0+1} \frac{f(\varepsilon_a)-f(\varepsilon_e)}{2\varepsilon_0 q_0}. \quad (15)$$

we note that in the derivation of (15) we used the inequality $|df/d\varepsilon| \gg f(\varepsilon)/\varepsilon$, which is the consequence of the condition formulated above on the degree of non-monochromaticity of the beam: $\delta\varepsilon/\varepsilon \ll 1$.

The difference between the values of the function $f(\varepsilon)$ f at the points $\varepsilon_a$ and $\varepsilon_e$ using formulas (14), is

$$f(\varepsilon_e) - f(\varepsilon_a) \approx 2\Delta\varepsilon \frac{df}{d\varepsilon} \approx \frac{\omega}{\gamma^2 m^2}\left(\frac{\varepsilon}{\delta\varepsilon}\right)^2, \quad \delta\varepsilon \gg \omega \quad (16)$$

(the derivative $df/d\varepsilon$ is calculated at the point $\varepsilon = \varepsilon_0$). Substituting (16) in (15) and using Eq. (9) for w, we obtain

$$\Delta w = \frac{\pi(eA_2)^2 K^2}{8\gamma m(1+K^2/2)}\left(\frac{\varepsilon}{\delta\varepsilon}\right)^2 \left[J_{n_0+1}(\alpha_e) \pm J_{n_0}(\alpha_e)\right]^2. \quad (17)$$

The difference $\Delta w$ of the total probabilities of the emission and absorption per unit time of a photon with frequency w determines the gain G at this frequency:

$$G = \frac{\Delta w \omega N_e}{E_2^2/8\pi}, \quad (18)$$

where $N_e$ is the electron density in the beam; $E_2$ is the amplitude of the electric field intensity in the amplified wave.

Substituting (17) in the definition (18) of G, we obtain for the gain per pass the expression

$$G = \frac{\Delta w \omega N_e}{E_2^2/8\pi} = \frac{\pi^2 e^2 N_e L K^2}{\gamma m \omega(1+K^2/2)}\left(\frac{\varepsilon}{\delta\varepsilon}\right)^2 \left[J_{n_0+1}(\alpha_e) \pm J_{n_0}(\alpha_e)\right]^2, \quad (19)$$

where L is the longitudinal dimension of the interaction region (in our formulation, L coincides with the undulator length).

## 4. CONCLUSION

The analysis proposed in this paper is valid when the role of the undulator is played by a traveling electromagnetic pump wave with planar polarization [79]. In this case, when the electron interacts with the wave it can capture many photons and emit one photon of the amplified wave. The corresponding gain differs from Eq. (19) only by a numerical factor.